\documentclass[journal]{IEEEtran}

\usepackage{amsmath,amssymb,mathtools}
\usepackage{bm}
\usepackage{cite}
\usepackage{algorithm}
\usepackage{algorithmic}
\usepackage{booktabs}
\usepackage{array}
\usepackage{graphicx}
\begin{document}

\title{Sharing Coefficient-Based Price Signals for Demand Response in Renewable Energy Communities}
\author{
Alireza~Shooshtari,~\IEEEmembership{Member,~IEEE,}
Antonio~Pepiciello,~\IEEEmembership{Member,~IEEE,}
and~José~Luis~Domínguez-García,~\IEEEmembership{Senior Member,~IEEE}%
\thanks{
Alireza Shooshtari, Antonio Pepiciello, and José Luis Domínguez-García are with the Catalonia Institute for Energy Research (IREC), Barcelona, Spain
(e-mail: ashooshtari@irec.cat; apepiciello@irec.cat; jldominguez@irec.cat).
The authors declare no conflicts of interest.
}
}
\maketitle
\thispagestyle{empty}

\begin{abstract}
Renewable energy communities can increase local photovoltaic (PV) use, but feeder-level surplus can still cause reverse power flow in low-voltage networks. Existing sharing coefficient methods are mainly used ex-post for surplus allocation and billing, so they do not directly guide demand toward hours and feeders where shared PV can reduce export. This paper proposes a sharing coefficient-based demand response framework that converts dynamic proportional allocation outcomes into household specific day-ahead price signals. The feeder-aware design first shares surplus within each feeder, while the feeder-agnostic design shares surplus through a single community pool. The energy community manager iteratively computes the allocation from submitted demand and PV forecasts, decomposes purchased energy into same-feeder, inter-feeder, and grid-import components, and coordinates household load reshaping through a convex optimization model. Using measured profiles from 15 households and AC power flow analysis, the framework reduces feeder reverse energy by 45.0\% and 44.6\% on selected high reverse energy days, and by 69.0\% and 66.3\% over the annual window, for the feeder-aware and feeder-agnostic cases, respectively. These results show that sharing coefficients can be used not only for ex-post billing, but also as operational price signals for demand response, with feeder-aware allocation providing an additional network benefit by accounting for household location in the low-voltage network.
\end{abstract}


\section{Introduction}

\IEEEPARstart{T}{he} transition toward low-carbon power systems is accelerating the decentralization of electricity generation, mainly driven by the increasing deployment of distributed energy resources such as residential photovoltaic (PV) systems. As a result, end users are evolving from passive consumers to active prosumers, and distribution networks are becoming active platforms for integrating local renewable energy.

In this context, energy communities have emerged as a framework to involve citizens and local actors in the energy transition. They enable participants to jointly organize local energy activities and share the resulting economic benefits \cite{shooshtari2025grid,caramizaru2020energy}. In Europe, this concept has been supported by the Clean Energy for all Europeans package \cite{ec2019clean} and formalized through Directive (EU) 2018/2001 and Directive (EU) 2019/944 \cite{eu2018red,eu2019iemd}. In residential PV-based communities, surplus generation can be shared among members through collective self-consumption schemes, improving local renewable energy utilization and reducing electricity imports and costs \cite{chaudhry2022renewable,gilmena2023analysis}.

Energy sharing is typically implemented through sharing coefficients, which define how surplus generation is allocated among participants. These coefficients can be fixed, proportional, priority-based, or dynamic. While they provide a structured mechanism for allocation and settlement, they do not directly influence consumption behavior, and therefore have limited impact on aligning demand with periods of local renewable generation \cite{ceer2025decentralised}. Demand response addresses this limitation by adjusting flexible loads in response to price signals \cite{vardakas2015survey,mohsenianrad2010optimal}. However, the effectiveness of such signals also depends on the distribution network, since the value of shared energy is influenced not only by its quantity but also by its location. Without considering network topology, economic benefits may not translate into grid benefits, motivating the design of price signals that account for both energy value and network impact \cite{gautier2025energy,johannsen2023energy}.

In recent years, many studies have investigated how prosumers can exchange energy inside local communities. Review papers have summarized different models for this purpose, including peer-to-peer trading, community self-consumption, local pricing, and prosumer decision-making \cite{zhou2023peer,gorbatcheva2024defining}. Energy collectives have also been proposed to coordinate community members while considering common objectives and fairness in the distribution of benefits \cite{moret2019energy}. Other studies have developed peer-to-peer market structures in which participants can trade energy directly according to their preferences \cite{sorin2019consensus}, or through bilateral contracts between distributed resources and flexible consumers \cite{morstyn2019bilateral}. These works provide important foundations for local energy exchange. However, their main focus is on the organization, pricing, and settlement of local transactions. The use of local sharing outcomes as price signals for household demand response and distribution network operation is not the central scope of these studies.

Another line of work has focused on sharing coefficients as a mechanism for allocating shared energy in renewable energy communities and collective self-consumption schemes. In the Portuguese regulatory context, fixed and time-variable sharing coefficients have been evaluated for renewable energy communities. The results show that time-variable coefficients can improve the use of shared energy, while the benefits received by each participant depend on their demand level and on whether self-consumption is considered before the sharing process \cite{queiroz2023assessment}. More recent studies have developed dynamic sharing coefficients to better capture the time-varying behavior of community members. These methods include consumption-proportional allocation, correlation-based allocation between consumption and surplus generation, trend-based allocation, and hybrid combinations of these coefficients \cite{gianaroli2024development}. Optimization-based approaches have also been proposed, where sharing coefficients are computed by a centralized optimization problem to minimize the total electricity bill of the community. Additional constraints can be included to improve the stability of the community and avoid unfair benefit allocation among members \cite{devillena2022financial}. A grid-informed sharing coefficient strategy has also been proposed to account for the feeder location of community members. This strategy can be applied to different fixed and dynamic sharing coefficients, including equal, proportional, and priority-based designs. In this approach, same-feeder sharing is prioritized before allocating the remaining surplus at the community level, allowing the sharing process to reflect the spatial structure of the distribution network \cite{shooshtari2025grid}. These studies improve the allocation of shared energy and can make participation in an energy community more economically attractive. However, their role is still mainly related to allocation and settlement.

Price-based demand response is a practical way to link energy sharing with flexible consumption. For example, in one energy-sharing model for PV prosumers, the internal price is defined based on the balance between local supply and demand, so each user can adjust their flexible load accordingly \cite{liu2017energysharing}. Later studies extend this idea by using home energy management systems to schedule appliances, storage, PV, and electric vehicles under dynamic peer-to-peer prices. These prices usually depend on the local supply–demand situation and the relation between buying and selling electricity from the grid \cite{alfaverh2023dynamic}. Some works also move toward fully decentralized approaches, where users make their own decisions based on time-of-use prices, incentives, and peer-to-peer trading, while keeping their data private \cite{hussain2024fully}. In renewable energy communities, centralized demand response has also been studied, where flexible demand is coordinated with local PV generation. This helps increase self-consumption and reduce overall electricity costs \cite{ercoli2025demand}. Although these studies show that price-based demand response can help shift consumption and better use local renewable energy, the price signals are usually based on overall supply–demand balance or market rules.

Network-aware local energy trading has also been studied to ensure that local exchanges do not create adverse impacts on the distribution grid. Existing approaches include sensitivity-based validation of peer-to-peer transactions under voltage and line constraints \cite{guerrero2019decentralized}, integration of peer-to-peer trading with distribution locational marginal pricing \cite{morstyn2020integrating}, distributed network-constrained local energy markets \cite{oliveira2023distributed}, and regional sharing prices or network usage fees for multi-region energy sharing \cite{chen2020peer}. 
While these approaches account for grid limitations, they are mostly based on market-clearing or network-optimization frameworks. 

Although local energy sharing, sharing coefficients, price-based demand response, and network-aware local markets have each been studied, they are still mostly treated separately. Sharing coefficients are commonly used as ex-post tools to allocate shared energy and settle benefits among members. Price-based demand response models can reshape household consumption, but their price signals are usually derived from aggregate demand–supply conditions, tariffs, incentives, or market rules, rather than from the shared energy actually allocated to each household. Network-aware market models include grid constraints, but mainly through market-clearing, locational pricing, or transaction-validation mechanisms. Therefore, to the best of our knowledge, existing studies have not used sharing coefficients that reflect both household-level shared-energy allocation and electrical location as household-specific price signals for demand response.

To address this gap, this paper extends our previous work in \cite{shooshtari2025grid} by moving the grid-informed sharing mechanism from ex-post surplus allocation and settlement to the design of household-specific day-ahead price signals for demand response. The proposed framework compares feeder-aware and feeder-agnostic allocation strategies, where the former prioritizes sharing among members connected to the same feeder and the latter allocates surplus through a single community pool. This enables the impact of allocation design to be evaluated in terms of demand shifting, economic performance, and grid operation.

The main contributions of this paper are summarized as follows:
\begin{itemize}
    \item A day-ahead demand response framework is proposed in which sharing coefficient allocation is translated into household-specific price signals for renewable energy communities.

    \item Feeder-aware and feeder-agnostic allocation strategies are incorporated and compared within the proposed demand response framework to study how allocation design affects local demand shifting and grid-impact indicators.

    \item An iterative coordination process between the energy community manager and households is formulated, where households respond to allocation-based price signals while preserving their daily energy consumption.

    \item The proposed framework is evaluated using measured residential PV and demand profiles mapped to a three-feeder low voltage network, considering both economic performance and grid operation metrics.
\end{itemize}

The remainder of this paper is organized as follows. Section II presents the proposed sharing coefficient-based demand response framework. Section III describes the case study, tariff parameters, and discusses the numerical results. Section IV concludes the paper.

\section{Methodology}

 This section presents the proposed day-ahead demand response framework. The method builds on the grid-informed sharing coefficient concept introduced in our previous work~\cite{shooshtari2025grid}, where sharing coefficients were used after the realization of production and consumption to allocate surplus energy and compute billing quantities. In the present work, this allocation principle is moved to the operational stage. The coefficients are computed from day-ahead PV generation and demand forecasts and are then used to construct household-specific price signals for demand response.

Among the sharing mechanisms evaluated in~\cite{shooshtari2025grid}, the dynamic proportional rule is selected because it allocates surplus according to the instantaneous unmet demand of the participants. It therefore adapts to the hourly mismatch between local production and consumption. The proposed framework applies this rule under two feeder-aware and feeder-agnostic allocation strategies. The feeder-aware strategy first allocates surplus within each feeder and then allocates any residual surplus at the community level. The feeder-agnostic strategy disregards feeder membership and allocates the community surplus in a single stage.

\subsection{Day-Ahead Information Exchange and Notation}

Indices $i$ and $j$ refer to energy community members, $f$ to feeders, and $t$ to hourly time periods. The allocation strategy is indexed by $m\in\{\mathrm{FA},\mathrm{AG}\}$, where $\mathrm{FA}$ and $\mathrm{AG}$ denote the feeder-aware and feeder-agnostic cases, respectively. Let $\mathcal{N}$ be the set of community members and let 
$\mathcal{T}=\{1,\ldots,T\}$ be the day-ahead scheduling horizon. 
The set of feeders is denoted by $\mathcal{F}$. Each member 
$i\in\mathcal{N}$ is connected to one feeder, denoted by 
$\phi_i\in\mathcal{F}$. For each feeder $f\in\mathcal{F}$, the subset 
of members connected to that feeder is defined as
\begin{equation}
    \mathcal{N}_f=\{i\in\mathcal{N}:\phi_i=f\}.
    \label{eq:feeder_member_set}
\end{equation}

All energy quantities are expressed per time interval. With hourly data, they are numerically equivalent to average power in kW over the hour. The forecasted PV generation of member $i$ at time $t$ is denoted by $p_{i,t}$. The baseline day-ahead demand forecast, before applying the proposed sharing coefficient-based demand response coordination, is denoted by $\ell_{i,t}$. The demand profile submitted to the energy community manager (ECM) during the coordination process is denoted by $x_{i,t}$. For a given submitted profile $x_{i,t}$, the self-consumed energy, the residual demand not covered by the member's own PV generation, and the PV surplus remaining after self-consumption are defined respectively as

\begin{equation}
    e_{i,t}^{\mathrm{self}}=\min\{x_{i,t},p_{i,t}\},
    \label{eq:self_consumption}
\end{equation}
\begin{equation}
    d_{i,t}=\max\{x_{i,t}-p_{i,t},0\},
    \label{eq:residual_demand}
\end{equation}
\begin{equation}
    s_{i,t}=\max\{p_{i,t}-x_{i,t},0\}.
    \label{eq:surplus}
\end{equation}

\subsection{Dynamic Proportional Sharing Coefficients}

For any subset of members $\mathcal{M}\subseteq\mathcal{N}$, let $i$ denote the member for which the coefficient is computed, and let $j$ be the summation index over all members in $\mathcal{M}$. The dynamic proportional demand coefficient of member $i\in\mathcal{M}$ is
\begin{equation}
    \omega_{i,t}(\mathcal{M}) =
    \begin{cases}
    \dfrac{d_{i,t}}{\sum_{j\in\mathcal{M}}d_{j,t}},
        & \sum_{j\in\mathcal{M}}d_{j,t}>0,\\[2mm]
    0, & \text{otherwise}.
    \end{cases}
    \label{eq:demand_coeff}
\end{equation}
Similarly, the dynamic proportional surplus coefficient of member $i\in\mathcal{M}$ is
\begin{equation}
    \nu_{i,t}(\mathcal{M}) =
    \begin{cases}
    \dfrac{s_{i,t}}{\sum_{j\in\mathcal{M}}s_{j,t}},
        & \sum_{j\in\mathcal{M}}s_{j,t}>0,\\[2mm]
    0, & \text{otherwise}.
    \end{cases}
    \label{eq:surplus_coeff}
\end{equation}
The coefficient in~\eqref{eq:demand_coeff} is an allocation key and not an energy quantity. In each allocation stage, the coefficient is multiplied by the available surplus, and the resulting amount is limited by the participant's residual demand. The coefficient in~\eqref{eq:surplus_coeff} is used to attribute the shared energy to surplus producers for settlement and price signal construction.

\subsection{Feeder-Agnostic Allocation}

The feeder-agnostic strategy allocates the community surplus in a single stage, without reflecting the location of the members within feeders. For member $i$, the provisional allocation is
\begin{equation}
    \tilde a_{i,t}
    =
    \omega_{i,t}(\mathcal{N})\sum_{j\in\mathcal{N}} s_{j,t}.
    \label{eq:agnostic_prealloc}
\end{equation}
Since the provisional allocation may exceed the residual demand of a member when the available surplus is larger than the total residual demand, the actual feeder-agnostic allocation is capped by the member's residual demand as
\begin{equation}
    a_{i,t}
    =
    \min\left\{d_{i,t},\tilde a_{i,t}\right\}.
    \label{eq:agnostic_alloc}
\end{equation}
The total allocation in the feeder-agnostic pool is
\begin{equation}
    L_t
    =
    \sum_{i\in\mathcal{N}}a_{i,t}.
    \label{eq:agnostic_total_alloc}
\end{equation}
The part of the feeder-agnostic allocation attributed to producer $i$ is
\begin{equation}
    b_{i,t}
    =
    \begin{cases}
    \dfrac{s_{i,t}}{\sum_{j\in\mathcal{N}} s_{j,t}}L_t,
    & \sum_{j\in\mathcal{N}} s_{j,t}>0,\\[2mm]
    0, & \text{otherwise}.
    \end{cases}
    \label{eq:agnostic_sold}
\end{equation}

Since the feeder-agnostic strategy allocates energy at the community level, the allocation received by a member is not explicitly separated into same-feeder and inter-feeder contributions. Therefore, the same-feeder part is estimated by comparing the shared energy attributed to surplus producers in feeder $f$ with the shared energy received by consumers in the same feeder. The fraction of the allocation received in feeder $f$ that can be treated as same-feeder sharing is

\begin{equation}
    \theta_{f,t}
    =
    \begin{cases}
    \min\left\{1,\dfrac{\sum_{i\in\mathcal{N}_f} b_{i,t}}{\sum_{i\in\mathcal{N}_f}a_{i,t}}\right\},
    &\sum_{i\in\mathcal{N}_f}a_{i,t}>0,\\[2mm]
    0, & \text{otherwise}.
    \end{cases}
    \label{eq:theta_ag}
\end{equation}
The feeder-agnostic allocation is then decomposed as
\begin{equation}
    A_{i,t}^{\mathrm{same,AG}}
    =
    \theta_{\phi_i,t}
    a_{i,t},
    \label{eq:ag_same}
\end{equation}
\begin{equation}
    A_{i,t}^{\mathrm{other,AG}}
    =
    \left(1-\theta_{\phi_i,t}\right)
    a_{i,t},
    \label{eq:ag_other}
\end{equation}
\begin{equation}
    G_{i,t}^{\mathrm{AG}}
    =
    d_{i,t}-a_{i,t}.
    \label{eq:ag_grid}
\end{equation}

\subsection{Feeder-Aware Allocation}

The feeder-aware allocation is carried out in two stages. The first stage uses the surplus available inside each feeder to supply the residual demand of members connected to that feeder. For member $i$, the provisional local allocation, before capping it by the residual demand of that member, is
\begin{equation}
    \tilde a_{i,t}^{\mathrm{loc}}
    =
    \omega_{i,t}(\mathcal{N}_{\phi_i})
    \sum_{j\in\mathcal{N}_{\phi_i}} s_{j,t}.
    \label{eq:local_prealloc}
\end{equation}
The actual local allocation is then obtained as
\begin{equation}
    a_{i,t}^{\mathrm{loc}}
    =
    \min\left\{
    d_{i,t},\tilde a_{i,t}^{\mathrm{loc}}
    \right\}.
    \label{eq:local_alloc}
\end{equation}
The total local allocation in feeder $f$ is
\begin{equation}
    L_{f,t}^{\mathrm{loc}}
    =
    \sum_{i\in\mathcal{N}_f}a_{i,t}^{\mathrm{loc}}.
    \label{eq:local_total_alloc}
\end{equation}
Since this stage only allocates surplus within the same feeder, $a_{i,t}^{\mathrm{loc}}$ is treated as same-feeder sharing. The demand and surplus not settled locally are carried to the second stage. These residual quantities are
\begin{equation}
    \bar d_{i,t}
    =
    d_{i,t}-a_{i,t}^{\mathrm{loc}},
    \label{eq:res_demand_fa}
\end{equation}
\begin{equation}
    \bar s_{i,t}
    =
    s_{i,t}
    -
    \nu_{i,t}(\mathcal{N}_{\phi_i})
    L_{\phi_i,t}^{\mathrm{loc}}.
    \label{eq:res_surplus_fa}
\end{equation}

The second stage allocates the remaining surplus at the community level according to the residual demand. The residual demand coefficient of member $i$ is
\begin{equation}
    \bar\omega_{i,t}
    =
    \begin{cases}
    \dfrac{\bar d_{i,t}}{\sum_{j\in\mathcal{N}}\bar d_{j,t}},
    & \sum_{j\in\mathcal{N}}\bar d_{j,t}>0,\\[2mm]
    0, & \text{otherwise}.
    \end{cases}
    \label{eq:residual_coeff_fa}
\end{equation}
The provisional community-level allocation, before capping it by the residual demand of the member, is
\begin{equation}
    \tilde a_{i,t}^{\mathrm{g}}
    =
    \bar\omega_{i,t}\sum_{j\in\mathcal{N}} \bar s_{j,t}.
    \label{eq:global_prealloc_fa}
\end{equation}
The actual community-level allocation is then
\begin{equation}
    a_{i,t}^{\mathrm{g}}
    =
    \min\left\{\bar d_{i,t},\tilde a_{i,t}^{\mathrm{g}}\right\}.
    \label{eq:global_alloc_fa}
\end{equation}
The total allocation in the community-level stage is
\begin{equation}
    L_t^{\mathrm{g}}
    =
    \sum_{i\in\mathcal{N}}a_{i,t}^{\mathrm{g}}.
    \label{eq:global_total_alloc_fa}
\end{equation}

The classification of the second-stage allocation follows from the residual structure created by the local stage. If the total surplus in a feeder is less than or equal to the total demand in that feeder, all local surplus is used in the first stage and only residual demand can remain. If the total surplus in a feeder is larger than the total demand in that feeder, all local demand is supplied in the first stage and only residual surplus can remain. Therefore, after the local stage, a feeder cannot contain both residual demand and residual surplus. Any allocation received in the community-level stage must consequently be supplied by surplus remaining in other feeders and is treated as inter-feeder sharing.
\begin{equation}
    A_{i,t}^{\mathrm{same,FA}}
    =
    a_{i,t}^{\mathrm{loc}}.
    \label{eq:fa_same}
\end{equation}
\begin{equation}
    A_{i,t}^{\mathrm{other,FA}}
    =
    a_{i,t}^{\mathrm{g}}.
    \label{eq:fa_other}
\end{equation}
\begin{equation}
    G_{i,t}^{\mathrm{FA}}
    =
    d_{i,t}
    -
    a_{i,t}^{\mathrm{loc}}
    -
    a_{i,t}^{\mathrm{g}}.
    \label{eq:fa_grid}
\end{equation}
Here, $A_{i,t}^{\mathrm{same,FA}}$ is the part treated as same-feeder sharing, $A_{i,t}^{\mathrm{other,FA}}$ is the part treated as inter-feeder sharing, and $G_{i,t}^{\mathrm{FA}}$ is the remaining grid import.

For either strategy $m\in\{\mathrm{FA},\mathrm{AG}\}$, the submitted demand is decomposed into self-consumption, same-feeder sharing, inter-feeder sharing, and grid import:
\begin{equation}
    x_{i,t}
    =
    e_{i,t}^{\mathrm{self}}
    +A_{i,t}^{\mathrm{same},m}
    +A_{i,t}^{\mathrm{other},m}
    +G_{i,t}^{m}.
    \label{eq:energy_balance_decomposition}
\end{equation}

\subsection{Sharing Coefficient-Based Price Signal}

The proposed demand response signal is derived from the allocation decomposition in~\eqref{eq:fa_same}--\eqref{eq:fa_grid} or~\eqref{eq:ag_same}--\eqref{eq:ag_grid}. The retail tariff is decomposed into an energy component $c^{\mathrm{en}}$, a network component $c^{\mathrm{net}}$, and other regulated charges $c^{\mathrm{reg}}$. Therefore, the price of importing energy from the grid is
\begin{equation}
    c^{\mathrm{grid}}
    =
    c^{\mathrm{en}}+c^{\mathrm{net}}+c^{\mathrm{reg}}.
    \label{eq:c_grid}
\end{equation}
Energy bought inside the community receives a discount on the network component. The discount is larger when the allocation is treated as same-feeder sharing than when it is treated as inter-feeder sharing. Let $\rho^{\mathrm{same}}$ and $\rho^{\mathrm{other}}$ be the fractions of the network component that remain in the same-feeder and inter-feeder prices, respectively, with $0\leq\rho^{\mathrm{same}}\leq\rho^{\mathrm{other}}\leq 1$. These fractions are exogenous tariff-design parameters, determined by the applicable regulation or by the community
contract rather than by the demand response optimization. The corresponding community purchase prices are
\begin{align}
    c^{\mathrm{same}}
        &= c^{\mathrm{en}}+\rho^{\mathrm{same}}c^{\mathrm{net}}+c^{\mathrm{reg}}, \label{eq:c_same}\\
    c^{\mathrm{other}}
        &= c^{\mathrm{en}}+\rho^{\mathrm{other}}c^{\mathrm{net}}+c^{\mathrm{reg}}. \label{eq:c_other}
\end{align}
Under this convention, $1-\rho^{\mathrm{same}}$ and $1-\rho^{\mathrm{other}}$ represent the network charge discounts applied to same-feeder and inter-feeder sharing, respectively.

For strategy $m\in\{\mathrm{FA},\mathrm{AG}\}$, the purchase cost associated with member $i$ at time $t$ is
\begin{equation}
C_{i,t}^{m}
=
c^{\mathrm{same}}A_{i,t}^{\mathrm{same},m}
+c^{\mathrm{other}}A_{i,t}^{\mathrm{other},m}
+c^{\mathrm{grid}}G_{i,t}^{m}.
\label{eq:purchase_cost}
\end{equation}
The total purchased energy after self-consumption is
\begin{equation}
q_{i,t}^{m}
=
A_{i,t}^{\mathrm{same},m}
+A_{i,t}^{\mathrm{other},m}
+G_{i,t}^{m}
=
d_{i,t}.
\label{eq:total_purchased_energy}
\end{equation}
For $q_{i,t}^{m}>0$, the household-specific price signal is defined as the effective unit cost of purchased energy implied by the allocation outcome:
\begin{equation}
\pi_{i,t}^{m}
=
\frac{
c^{\mathrm{same}}A_{i,t}^{\mathrm{same},m}
+c^{\mathrm{other}}A_{i,t}^{\mathrm{other},m}
+c^{\mathrm{grid}}G_{i,t}^{m}
}{
q_{i,t}^{m}
}.
\label{eq:price_signal}
\end{equation}

If $q_{i,t}^{m}=0$, member $i$ does not purchase energy in that hour and the ratio in~\eqref{eq:price_signal} is not directly defined. This case does not affect the household objective as long as the scheduled demand remains below or equal to the PV generation, because the purchased energy term is then zero. To obtain a price signal for a possible upward shift above $p_{i,t}$, the ECM evaluates the same effective-price rule after increasing only the demand of member $i$ in that hour to $p_{i,t}+\delta^{\mathrm{p}}$, while keeping all other submitted quantities unchanged. The perturbation $\delta^{\mathrm{p}}$ is used only for this price evaluation and is set to $10^{-3}$~kWh in the numerical study, which is small relative to the hourly household demand values and does not affect the final allocation or billing calculation.

\subsection{Household Demand Response Optimization}

The household response model follows the general structure of an aggregate 
load shifting formulation previously used for price-based energy sharing, 
where an internal community price is derived from the supply-demand ratio 
of the sharing zone~\cite{liu2017energysharing}. In contrast, the present 
work derives household and hour-specific prices from the allocation 
decomposition into same-feeder sharing, inter-feeder sharing, and grid 
import. The response is modeled at the household level by allowing the 
day-ahead demand profile to be reshaped within the observed daily operating 
range of that household.

For a given price vector $\bm{\pi}_i^{m}=\{\pi_{i,t}^{m}\}_{t\in\mathcal{T}}$,
household $i$ solves
\begin{subequations}
\label{eq:household_problem}
\begin{align}
    \min_{\{y_{i,t},\, z_{i,t}\}_{t\in\mathcal{T}}}\quad
    &\sum_{t\in\mathcal{T}}
    \left[
    \pi_{i,t}^{m}\, z_{i,t}
    +\frac{\mu_i}{2}\left(y_{i,t}-\ell_{i,t}\right)^2
    \right] \label{eq:obj}\\
    \mathrm{s.t.}\quad
    &\sum_{t\in\mathcal{T}} y_{i,t}
    =
    \sum_{t\in\mathcal{T}} \ell_{i,t}, \label{eq:energy_neutrality}\\
    &y_{i}^{\mathrm{min}}\leq y_{i,t}\leq y_{i}^{\mathrm{max}},
    \quad \forall t\in\mathcal{T}, \label{eq:bounds}\\
    &z_{i,t} \geq y_{i,t} - p_{i,t},
    \quad \forall t\in\mathcal{T}, \label{eq:epigraph_purchase}\\
    &z_{i,t} \geq 0,
    \quad \forall t\in\mathcal{T}. \label{eq:epigraph_nonneg}
\end{align}
\end{subequations}
The objective~\eqref{eq:obj} contains two terms. The first charges the 
purchased energy $z_{i,t}$ at the household-specific price $\pi_{i,t}^{m}$,
where constraints~\eqref{eq:epigraph_purchase}--\eqref{eq:epigraph_nonneg} 
ensure that $z_{i,t}$ equals the energy drawn from outside the household's 
own PV generation. The second term penalizes deviations from the base 
day-ahead demand forecast, with parameter $\mu_i>0$ controlling the 
household's willingness to shift consumption.

The equality constraint~\eqref{eq:energy_neutrality} preserves the total 
daily consumption of the household. Constraint~\eqref{eq:bounds} represents 
the flexibility range of the aggregate demand. The lower and upper bounds are 
defined from the base day-ahead demand profile of the same household as
\begin{equation}
    y_{i}^{\mathrm{min}}
    =
    \min_{\tau\in\mathcal{T}}\ell_{i,\tau},
    \label{eq:lower_bound}
\end{equation}
\begin{equation}
    y_{i}^{\mathrm{max}}
    =
    \max_{\tau\in\mathcal{T}}\ell_{i,\tau},
    \label{eq:upper_bound}
\end{equation}
for all $t\in\mathcal{T}$. Therefore, the optimized profile can shift 
consumption across hours while remaining within the household's base 
daily demand range.

For a fixed price vector, problem~\eqref{eq:household_problem} is solved 
independently by each household. The resulting direct response is denoted 
by $y_{i,t}^{m}$. In the iterative coordination procedure, this response 
is used by the ECM to update the submitted demand profile.

\subsection{Iterative ECM-Household Coordination}

The proposed price signal is endogenous because it is computed from the sharing allocation. However, the sharing allocation depends on the demand profiles submitted by the households. Therefore, a price signal computed only once from the base demand forecast may become inconsistent after households shift their demand in response to that price. Once the demand profiles change, the residual demand, surplus allocation, grid import, and consequently the price signal also change. For this reason, an iterative coordination procedure is used to obtain a stable day-ahead schedule in which the final prices and the final household demand profiles are mutually consistent.
The procedure is carried out separately for the feeder-aware and feeder-agnostic strategies.

For a selected strategy $m\in\{\mathrm{FA},\mathrm{AG}\}$, the ECM 
initializes the submitted demand profile with the base forecast, 
$x_{i,t}^{(0)}=\ell_{i,t}$. At iteration $k$, the ECM applies the 
corresponding sharing allocation to the current submitted demand 
$x_{i,t}^{(k)}$ and the PV forecast $p_{i,t}$. This gives the 
self-consumed energy, same-feeder allocation, inter-feeder allocation, 
and grid import for each household and time period. The ECM then converts 
this decomposition into household-specific prices $\pi_{i,t}^{m,(k)}$ 
using~\eqref{eq:price_signal}.

After receiving the price vector, each household 
solves~\eqref{eq:household_problem} using $\bm{\pi}_i^{m,(k)}$ and returns 
its direct response $y_{i,t}^{m,(k+1)}$. To avoid abrupt changes between two successive allocations and oscillations in the iterative procedure, the ECM updates the submitted demand profile as
\begin{equation}
    x_{i,t}^{(k+1)}
    =
    \alpha y_{i,t}^{m,(k+1)}
    +(1-\alpha)x_{i,t}^{(k)},
    \quad 0<\alpha\leq 1
    \label{eq:damping_update}
\end{equation}
Equation~\eqref{eq:damping_update} introduces a relaxation step in the 
fixed-point coordination between allocation-based prices and household 
responses. Instead of replacing the submitted profile directly by the 
household response, the ECM blends the new response with the previous 
submitted profile. The 
parameter $\alpha$ is used only in the ECM update and does not modify the 
household feasible set in~\eqref{eq:household_problem}.
The sharing allocation and the corresponding prices are then recomputed 
from $x_{i,t}^{(k+1)}$.

The iteration is monitored using normalized changes in both the price 
signal and the submitted demand profile:
\begin{equation}
    \Delta_{\pi}^{(k+1)}
    =
    \max_{i,t}
    \frac{
    \left|\pi_{i,t}^{m,(k+1)}-\pi_{i,t}^{m,(k)}\right|
    }{
    c^{\mathrm{grid}}
    },
    \label{eq:price_delta_norm}
\end{equation}
\begin{equation}
    \Delta_{x}^{(k+1)}
    =
    \max_{i,t}
    \frac{
    \left|x_{i,t}^{(k+1)}-x_{i,t}^{(k)}\right|
    }{
    \max_{\tau\in\mathcal{T}}\ell_{i,\tau}
    }.
    \label{eq:load_delta_norm}
\end{equation}
The iteration stops when
\begin{equation}
    \max\left\{\Delta_{\pi}^{(k+1)},\Delta_x^{(k+1)}\right\}
    \leq \varepsilon,
    \label{eq:joint_convergence}
\end{equation}

or when the maximum number of iterations is reached. The final submitted 
profile $x_{i,t}^{(k^\star)}$ is taken as the day-ahead demand schedule 
induced by strategy $m$. The resulting feeder-aware and feeder-agnostic 
schedules are then compared with the base forecast using the economic 
and network performance metrics reported in the results section. The 
complete procedure is summarized in Algorithm~\ref{alg:ecm}.

\begin{algorithm}[!bt]
\caption{Sharing Coefficient-Based Demand Response Coordination}
\label{alg:ecm}
\begin{algorithmic}[1]
\STATE Select allocation strategy $m\in\{\mathrm{FA},\mathrm{AG}\}$.
\STATE Initialize $x_{i,t}^{(0)}=\ell_{i,t}$ for all $i\in\mathcal{N}$ and $t\in\mathcal{T}$.
\FOR{$k=0,1,\ldots,K_{\max}-1$}
    \STATE ECM computes self-consumption, residual demand, and surplus from the current profiles $x_{i,t}^{(k)}$ and PV forecasts $p_{i,t}$.
    \STATE ECM applies the selected sharing allocation strategy $m$ and obtains $A_{i,t}^{\mathrm{same},m,(k)}$, $A_{i,t}^{\mathrm{other},m,(k)}$, and $G_{i,t}^{m,(k)}$.
    \STATE ECM computes household-specific price signals $\pi_{i,t}^{m,(k)}$ using~\eqref{eq:price_signal}.
    \STATE Each household solves~\eqref{eq:household_problem} and returns its direct response $y_{i,t}^{m,(k+1)}$.
    \STATE ECM updates the submitted profile using the relaxation step in~\eqref{eq:damping_update}.
    \STATE ECM recomputes the allocation and prices from $x_{i,t}^{(k+1)}$ and evaluates $\Delta_{\pi}^{(k+1)}$ and $\Delta_x^{(k+1)}$.
    \IF{$\max\{\Delta_{\pi}^{(k+1)},\Delta_x^{(k+1)}\}\leq\varepsilon$}
        \STATE Stop and set $k^\star=k+1$.
    \ENDIF
\ENDFOR
\STATE If the stopping criterion is not met, set $k^\star=K_{\max}$.
\STATE Return the final day-ahead schedule $x_{i,t}^{(k^\star)}$. 
\end{algorithmic}
\end{algorithm}

\section{Case Study and Results}

\subsection{Case Study Description}

The objective of this case study is to evaluate the proposed sharing coefficient-based demand response framework under realistic residential conditions, comparing the feeder-aware and feeder-agnostic allocation strategies in terms of economic performance and grid operation metrics. The numerical study is based on the Irish energy-community dataset in~\cite{trivedi2024dataset}, which provides measured hourly residential consumption and PV generation profiles for 2020. In this study, the measured profiles are used as the day-ahead demand and PV forecasts submitted to the energy community manager.

Fifteen households are considered over the common data window from January~1, 2020 at 01:00 to December~11, 2020 at 23:00. The selected households include members with and without local PV generation. They are mapped to three radial low voltage feeders, with five households connected to each feeder. The resulting test community is summarized in Table~\ref{tab:community_composition}.

\begin{table}[!t]
\centering
\caption{Test Community Used in the Numerical Study}
\label{tab:community_composition}
\scriptsize
\begin{tabular}{c l c c c c c}
\toprule
Feeder & Members & Prosumers & Consumers & Load & PV & PV/load \\
 & & & & (MWh) & (MWh) & (\%) \\
\midrule
F1 & H1--H5 & 5 & 0 & 21.89 & 9.28 & 42.4 \\
F2 & H6--H10 & 2 & 3 & 28.95 & 3.71 & 12.8 \\
F3 & H11--H15 & 2 & 3 & 21.52 & 3.53 & 16.4 \\
\midrule
Total & H1--H15 & 9 & 6 & 72.36 & 16.51 & 22.8 \\
\bottomrule
\end{tabular}
\end{table}

The mapping gives a spatially non-uniform distribution of PV generation across the three feeders. Feeder F1 has the highest PV to load ratio, while F2 and F3 have lower PV penetration.

The low voltage network is modeled in Pandapower~\cite{thurner2018pandapower} and evaluated using AC power flow. The network consists of a 20kV upstream grid and three radial 0.4kV feeders. Each feeder is connected to the upstream grid through a 20/0.4kV, 25kVA transformer and supplies five household buses. Loads are modeled with a 0.95 lagging power factor, and PV units are modeled at unity power factor.

The retail tariff follows the decomposition in~\eqref{eq:c_grid}--\eqref{eq:c_other}. The numerical values are adopted from our previous sharing coefficient study~\cite{shooshtari2025grid} and are reported in Table~\ref{tab:tariff_parameters}.

\begin{table}[!t]
\centering
\caption{Tariff Parameters Used in the Numerical Study}
\label{tab:tariff_parameters}
\scriptsize
\begin{tabular}{l c c}
\toprule
Parameter & Value & Unit \\
\midrule
$c^{\mathrm{en}}$ & 0.1137 & EUR/kWh \\
$c^{\mathrm{net}}$ & 0.1092 & EUR/kWh \\
$c^{\mathrm{reg}}$ & 0.0333 & EUR/kWh \\
$c^{\mathrm{grid}}$ & 0.2562 & EUR/kWh \\
$\rho^{\mathrm{same}}$ & 0.2 & -- \\
$\rho^{\mathrm{other}}$ & 0.8 & -- \\
$c^{\mathrm{same}}$ & 0.1688 & EUR/kWh \\
$c^{\mathrm{other}}$ & 0.2344 & EUR/kWh \\
\bottomrule
\end{tabular}
\end{table}

The demand response parameters are kept fixed across all numerical comparisons. The discomfort coefficient is set to $\mu_i=0.2$~EUR/kWh$^2$ for all households, and the ECM relaxation factor is set to $\alpha=0.5$. The convergence tolerance is set to $\varepsilon=10^{-4}$, and the maximum number of coordination iterations is set to $K_{\max}=50$. The tolerance is applied to the normalized price and demand residuals in~\eqref{eq:price_delta_norm}--\eqref{eq:load_delta_norm}. With the tariff values in Table~\ref{tab:tariff_parameters}, this corresponds to a maximum price update below $2.56\times10^{-5}$~EUR/kWh and a maximum demand update below $10^{-4}$ of the household demand scale. The choices of $\mu_i$, $\alpha$, and $K_{\max}$ are justified through sensitivity analyses and convergence statistics in the operational and economic performance section.

Three cases are compared. The base day-ahead demand forecast without demand response is used as the baseline. The proposed coordination framework is then evaluated using the feeder-aware allocation strategy and the feeder-agnostic allocation strategy. The same input profiles, tariff parameters, demand response parameters, and network model are used in all cases.

\subsection{Performance Indicators}
The main network indicator used in this study is the reverse power flow in each low voltage feeder. This metric is directly related to the objective of the proposed demand response framework, which is to shift demand toward periods of local PV availability and reduce the surplus power exported from feeders. High feeder reverse power can increase upstream power exchange and limit the ability of low voltage feeders to host additional distributed PV generation. In the AC power-flow results, feeder reverse power is taken as the exported active power at the feeder head. The reported reverse energy is obtained by integrating this feeder export over the evaluation period.

In addition to reverse energy, the results report the peak total reverse power, the peak import from the upstream grid, the main-grid export energy, network energy losses, and the minimum bus voltage obtained from the AC power-flow simulations. These indicators are used to verify that reverse energy reduction does not introduce adverse effects on other operating quantities.

The economic impact is evaluated using the total community bill. The relative saving with respect to the base demand profile without demand response is
\begin{equation}
    \eta^{\mathrm{save}}
    =
    \frac{B^{\mathrm{base}}-B}{B^{\mathrm{base}}}
    \times 100,
    \label{eq:bill_saving}
\end{equation}
where $B^{\mathrm{base}}$ is the baseline bill and $B$ is the bill obtained under the evaluated demand response strategy.

\subsection{Operational and Economic Performance}

The operational performance is first evaluated on days with pronounced feeder-level reverse power, since these are the conditions under which the proposed demand response framework is most relevant. Such days indicate large PV surplus relative to local demand and are therefore the most informative cases for assessing whether allocation-based price signals can shift demand toward periods of local generation and reduce feeder export. From the baseline case, the days in which all three feeders experience reverse energy are identified, and the five days with the largest total feeder reverse energy are selected. The selected days are reported in Table~\ref{tab:high_reverse_days}. In the table, the total value is the sum of the daily reverse energy over the three feeders.

\begin{table}[!t]
\centering
\caption{High Reverse Energy Days Used for Performance Evaluation}\label{tab:high_reverse_days}
\scriptsize
\begin{tabular}{c c c c c}
\toprule
Date & Total & F1 & F2 & F3 \\
 & (kWh) & (kWh) & (kWh) & (kWh) \\
\midrule
2020-09-07 & 49.39 & 40.85 & 3.01 & 5.53 \\
2020-07-08 & 44.48 & 38.92 & 3.47 & 2.09 \\
2020-08-04 & 43.52 & 35.99 & 4.61 & 2.92 \\
2020-06-21 & 43.14 & 36.29 & 4.92 & 1.92 \\
2020-05-14 & 40.51 & 19.12 & 15.25 & 6.14 \\
\bottomrule
\end{tabular}
\end{table}

Table~\ref{tab:high_reverse_days} shows that reverse energy is not uniformly distributed across the feeders. In most selected days, F1 is the dominant exporting feeder, which is consistent with its higher PV to load ratio in Table~\ref{tab:community_composition}.

The average results over the selected days are summarized in Tables~\ref{tab:avg_high_reverse_main} and~\ref{tab:avg_high_reverse_support}. The feeder-aware strategy reduces the average reverse energy from 44.21 in the baseline to 24.32~kWh/day, corresponding to a 45\% reduction. The feeder-agnostic strategy gives a comparable reduction, reaching 24.51~kWh/day. The difference in total reverse energy is modest over this set of days, but the feeder-aware strategy also gives slightly lower main-grid export and community bill.

\begin{table}[!t]
\centering
\caption{Average Main Performance Over the Five High Reverse Energy Days}
\label{tab:avg_high_reverse_main}
\scriptsize
\setlength{\tabcolsep}{3pt}
\begin{tabular}{l c c c c c}
\toprule
Case & Rev. energy & Rev. peak & Main export & Bill & Saving \\
 & (kWh/day) & (kW) & (kWh/day) & (EUR/day) & (\%) \\
\midrule
Base & 44.21 & 7.99 & 35.34 & 31.57 & 0.00 \\
Feeder-aware & 24.32 & 4.72 & 9.45 & 24.95 & 21.18 \\
Feeder-agnostic & 24.51 & 4.74 & 10.01 & 25.23 & 20.19 \\
\bottomrule
\end{tabular}
\end{table}

\begin{table}[!t]
\centering
\caption{Average Supporting Indicators Over the Five High Reverse Energy Days}
\label{tab:avg_high_reverse_support}
\scriptsize
\begin{tabular}{l c c c c}
\toprule
Case & Peak import & Loss & $V_{\min}$ & Iteration \\
 & (kW) & (kWh/day) & (p.u.) & \\
\midrule
Base & 10.57 & 0.158 & 0.9955 & -- \\
Feeder-aware & 8.04 & 0.090 & 0.9963 & 15.6 \\
Feeder-agnostic & 8.04 & 0.091 & 0.9964 & 15.4 \\
\bottomrule
\end{tabular}
\end{table}

As shown in Table~\ref{tab:avg_high_reverse_support}, both demand response strategies reduce the peak import and network energy losses, while the minimum voltage remains close to the baseline value in all cases. The reduction in main-grid export is also substantial, decreasing from 35.34~kWh/day to 9.45~kWh/day with the feeder-aware strategy and to 10.01~kWh/day with the feeder-agnostic strategy. This indicates that the shifted demand absorbs a larger share of the available PV generation within the community. The coordination process converges in a similar number of iterations for both strategies, with average values of 15.6 and 15.4 iterations for the feeder-aware and feeder-agnostic cases, respectively.

To examine the mechanism behind the reverse power flow reduction, May~14, 2020 is analyzed in more detail. This day is selected as a representative high reverse energy case because the baseline reverse energy is not concentrated in a single feeder. As reported in Table~\ref{tab:high_reverse_days}, the baseline profile produces 19.12, 15.25, and 6.14~kWh of reverse energy in F1, F2, and F3, respectively.

\begin{figure}[!t]
    \centering
    \includegraphics[width=\columnwidth]{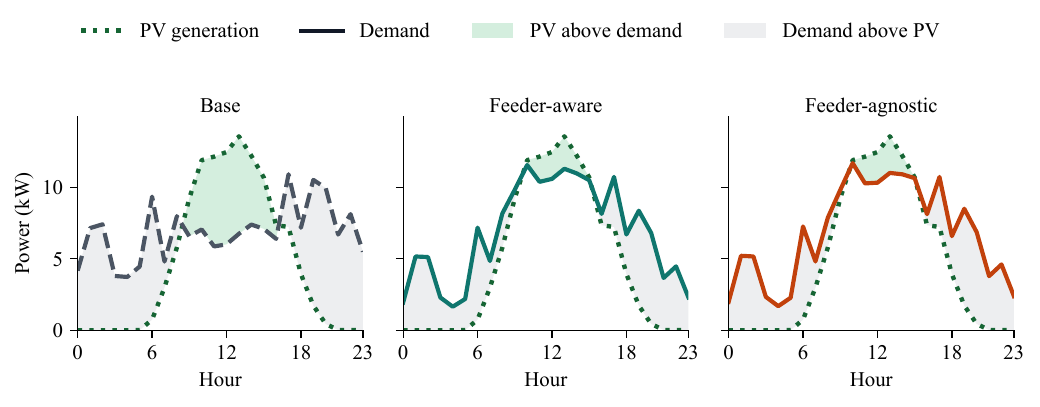}
    \caption{Community demand and PV generation profiles on May~14, 2020. The shaded regions indicate periods with PV surplus and residual demand.}
    \label{fig:demand_pv_20200514}
\end{figure}

Fig.~\ref{fig:demand_pv_20200514} shows the community demand and PV generation profiles under the baseline profile and under the two demand response strategies. In the baseline case, a large part of the PV generation occurs during hours in which the community demand is relatively low, leading to surplus generation around midday. After applying the allocation-based price signals, both demand response strategies shift part of the demand toward the PV production period. The feeder-aware and feeder-agnostic demand profiles are close at the aggregate community level, which indicates that both strategies produce a similar overall load shifting response.

\begin{figure}[!t]
    \centering
    \includegraphics[width=\columnwidth]{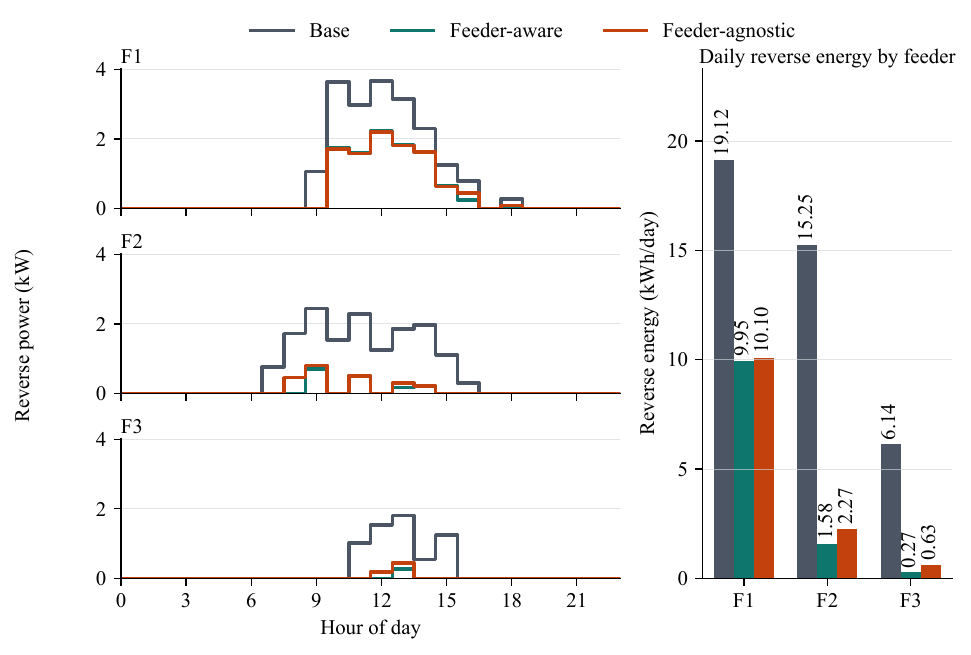}
    \caption{Hourly feeder reverse power and daily feeder reverse energy on May~14, 2020.}
    \label{fig:reverse_20200514}
\end{figure}

The corresponding feeder-level reverse energy profiles are shown in Fig.~\ref{fig:reverse_20200514}. The demand shift observed in Fig.~\ref{fig:demand_pv_20200514} reduces reverse power in all three feeders during the PV production hours. The feeder-aware and feeder-agnostic strategies produce similar reductions in F1, whereas the advantage of feeder-aware allocation is more visible in F2 and F3. This indicates that the feeder-aware signal mainly affects the spatial distribution of the response, rather than only the aggregate community demand.

Tables~\ref{tab:results_20200514_main} and~\ref{tab:results_20200514_support} report the corresponding operational and economic indicators. Relative to the baseline profile, the feeder-aware strategy reduces the total feeder reverse energy from 40.51 to 11.80~kWh, while the feeder-agnostic strategy reduces it to 13.00~kWh. The feeder-aware strategy also gives the lowest peak total reverse power, main-grid export energy, network energy loss, and community bill. These results show that the main improvement comes from the allocation-based demand response mechanism, while the feeder-aware allocation provides an additional network oriented benefit.

\begin{table}[!t]
\centering
\caption{Main Performance on May~14, 2020}
\label{tab:results_20200514_main}
\scriptsize
\begin{tabular}{l c c c c c}
\toprule
Case & Rev. energy & Rev. peak & Main export & Bill & Saving \\
 & (kWh) & (kW) & (kWh) & (EUR) & (\%) \\
\midrule
Base & 40.51 & 6.82 & 36.43 & 28.98 & 0.00 \\
Feeder-aware & 11.80 & 2.27 & 7.68 & 21.46 & 25.95 \\
Feeder-agnostic & 13.00 & 2.56 & 8.14 & 21.90 & 24.42 \\
\bottomrule
\end{tabular}
\end{table}

\begin{table}[!t]
\centering
\caption{Supporting Indicators on May~14, 2020}
\label{tab:results_20200514_support}
\scriptsize
\begin{tabular}{l c c c}
\toprule
Case & Peak import & Loss & $V_{\min}$ \\
 & (kW) & (kWh) & (p.u.) \\
\midrule
Base & 9.53 & 0.133 & 0.9969 \\
Feeder-aware & 6.65 & 0.065 & 0.9976 \\
Feeder-agnostic & 6.79 & 0.066 & 0.9976 \\
\bottomrule
\end{tabular}
\end{table}

The same parameter settings are then applied to the complete data window to verify whether the observed behavior persists beyond the selected high reverse energy days. The annual evaluation uses only complete 24-hour days within the common data window of the 15 households. Because the common window starts on January~1, 2020 at 01:00 and ends on December~11, 2020 at 23:00, January~1 is excluded. Therefore, the annual results are computed over 345 complete days, from January~2 to December~11, 2020. All 345 days are included in the annual indicators. When the stopping criterion is not reached before $K_{\max}$, the final submitted profile at $K_{\max}$ is used.

The annual network and economic results are reported in Tables~\ref{tab:annual_main_results} and~\ref{tab:annual_feeder_reverse}. The feeder-aware strategy reduces annual feeder reverse energy from 3618.4 to 1120.2~kWh, corresponding to a 69.0\% reduction. The feeder-agnostic strategy also gives a substantial reduction, reaching 1218.2~kWh, or 66.3\%. Thus, most of the reverse energy reduction is obtained by the allocation-based demand response mechanism itself, while the feeder-aware allocation provides an additional reduction by accounting for feeder location.

\begin{table}[!t]
\centering
\caption{Annual Network and Economic Performance}
\label{tab:annual_main_results}
\scriptsize
\begin{tabular}{l c c c c c}
\toprule
Case & Rev. energy & Main export & Loss & Bill & Saving \\
 & (kWh) & (kWh) & (kWh) & (EUR) & (\%) \\
\midrule
Base & 3618.4 & 2015.5 & 94.04 & 15898.59 & 0.00 \\
Feeder-aware & 1120.2 & 163.4 & 85.05 & 14846.75 & 6.62 \\
Feeder-agnostic & 1218.2 & 173.6 & 85.21 & 14884.73 & 6.38 \\
\bottomrule
\end{tabular}
\end{table}

The same trend is observed for main-grid export. The baseline profile exports 2015.5~kWh to the upstream grid over the annual window, whereas the feeder-aware and feeder-agnostic strategies reduce this value to 163.4 and 173.6~kWh, corresponding to reductions of 91.9\% and 91.4\%, respectively. Network energy losses are also reduced from 94.04~kWh to 85.05~kWh in the feeder-aware case and to 85.21~kWh in the feeder-agnostic case. The stopping criterion is reached for 330 of the 345 days in the feeder-aware case and for 343 days in the feeder-agnostic case.

Table~\ref{tab:annual_feeder_reverse} reports the annual reverse energy by feeder. In the baseline case, the reverse energy is dominated by F1, which is consistent with the high PV to load ratio of this feeder. Both demand response strategies reduce reverse energy in all feeders. The feeder-aware strategy leaves less reverse energy than the feeder-agnostic strategy in F1, F2, and F3, confirming that the benefit of feeder-aware allocation is consistent at the feeder level.

\begin{table}[!t]
\centering
\caption{Annual Feeder-Level Reverse Energy}
\label{tab:annual_feeder_reverse}
\scriptsize
\begin{tabular}{l c c c c}
\toprule
Case & F1 & F2 & F3 & Total \\
 & (kWh) & (kWh) & (kWh) & (kWh) \\
\midrule
Base & 3015.1 & 392.9 & 210.4 & 3618.4 \\
Feeder-aware & 1111.4 & 6.3 & 2.4 & 1120.2 \\
Feeder-agnostic & 1201.9 & 11.0 & 5.3 & 1218.2 \\
\bottomrule
\end{tabular}
\end{table}

The annual bill is reduced from 15898.59~EUR in the baseline case to 14846.75~EUR with the feeder-aware strategy and 14884.73~EUR with the feeder-agnostic strategy. These annual savings are lower than those observed on the selected high reverse energy days because the complete data window also includes days with lower PV production, limited surplus generation, or weaker reverse power flow conditions. On such days, there is less opportunity for the allocation-based price signal to shift demand toward locally available PV. Nevertheless, the feeder-aware strategy gives the lowest annual bill together with the lowest annual reverse energy and main-grid export.

Finally, the selected demand response and coordination parameters are justified through sensitivity analysis. The discomfort coefficient $\mu_i$ determines how strongly households penalize deviations from their baseline demand profiles, while the ECM update factor $\alpha$ controls the relaxation of the iterative coordination process. These two parameters are therefore examined separately.

For $\mu_i$, the sensitivity analysis reports both the reverse energy and the normalized demand deviation. The latter measures the relative change between the optimized demand profiles and the baseline profiles and is used only as a comparative indicator of demand reshaping, not as an absolute comfort threshold.

\begin{figure}[!tb]
    \centering
    \includegraphics[width=\columnwidth]{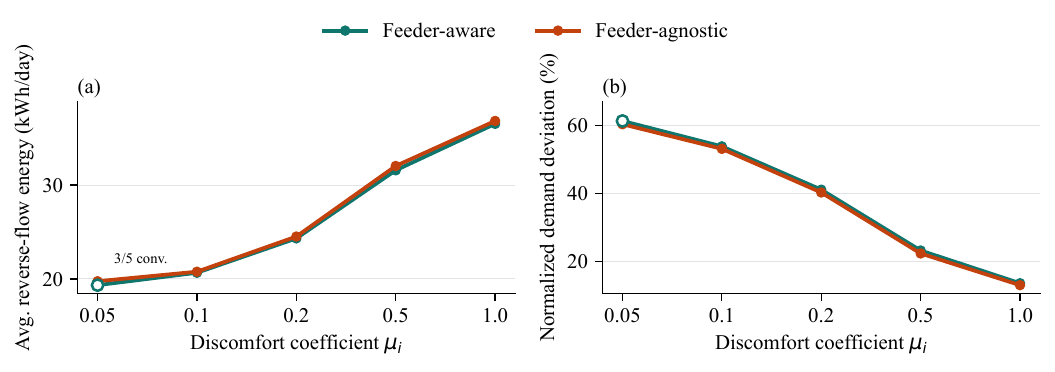}
    \caption{Sensitivity to the discomfort coefficient over the five high reverse energy days: (a) average feeder reverse energy and (b) normalized demand deviation.}
    \label{fig:sensitivity_mu_tradeoff}
\end{figure}

Fig.~\ref{fig:sensitivity_mu_tradeoff} shows the resulting trade-off. Smaller values of $\mu_i$ allow stronger demand reshaping and therefore lead to lower reverse energy. However, this improvement is obtained at the cost of larger deviations from the baseline demand profiles. For the feeder-aware strategy, reducing $\mu_i$ from 0.2 to 0.1 decreases the average reverse energy, but increases the normalized demand deviation from 41.08\% to 53.80\%. The lowest value, $\mu_i=0.05$, gives the strongest reverse energy reduction, but it also produces the largest demand deviation. In the feeder-aware case, convergence is reached for only three of the five selected days.

Larger values of $\mu_i$ lead to more conservative demand reshaping, but the reverse energy reduction becomes weaker. The value $\mu_i=0.2$~EUR/kWh$^2$ is therefore selected as a representative intermediate setting. It provides substantial reverse energy mitigation while avoiding the more aggressive demand reshaping and convergence issues observed for smaller discomfort coefficients. The feeder-aware strategy gives slightly lower reverse energy than the feeder-agnostic strategy for all tested values of $\mu_i$, although the difference remains moderate.

\begin{figure}[!tb]
    \centering
    \includegraphics[width=\columnwidth]{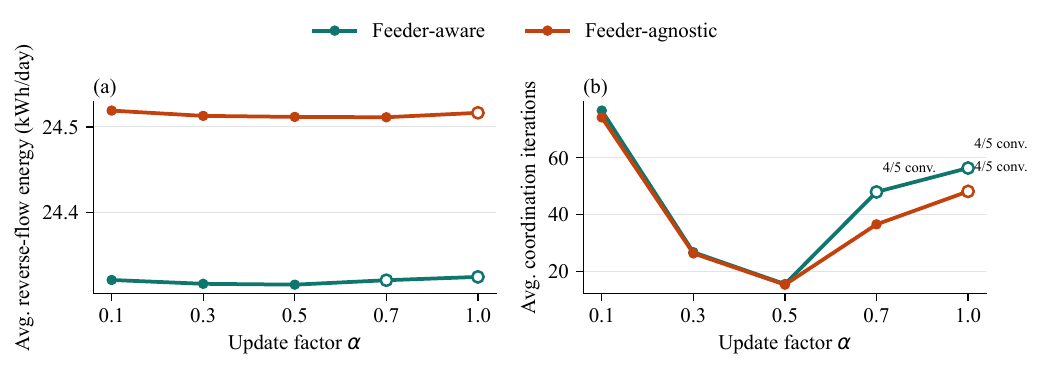}
    \caption{Sensitivity to the ECM update factor over the selected test days: (a) average feeder reverse energy and (b) average number of coordination iterations. Open markers indicate cases in which the stopping criterion is not reached for all tested days.}
    \label{fig:sensitivity_alpha}
\end{figure}

For $\alpha$, the sensitivity analysis is performed with $\mu_i=0.2$~EUR/kWh$^2$. Unlike $\mu_i$, the update factor does not change the household optimization problem. It only determines how much of the direct household response is incorporated into the submitted demand profile at each ECM iteration. Fig.~\ref{fig:sensitivity_alpha} shows that the average reverse energy is nearly unchanged across the tested values of $\alpha$, indicating that the final operating point is weakly affected by the update factor when the coordination process converges. The main effect of $\alpha$ is on convergence speed and robustness. Smaller values, such as $\alpha=0.1$ and $\alpha=0.3$, converge for all selected days but require more iterations. Larger values, such as $\alpha=0.7$ and $\alpha=1.0$, lead to partial non-convergence before $K_{\max}$ in some cases. Therefore, $\alpha=0.5$ is selected because it reaches the stopping criterion for all selected cases with a moderate number of iterations.

The maximum number of coordination iterations is used as a computational stopping safeguard. It is set to $K_{\max}=50$, which is above the largest number of iterations required by any converged annual case. In the annual evaluation, the maximum convergence iteration is 41 for the feeder-aware strategy and 29 for the feeder-agnostic strategy. For the remaining non-converged cases, a follow-up check with $K_{\max}=1000$ showed that the feeder-aware cases still did not satisfy the stopping criterion. Therefore, $K_{\max}=50$ is sufficient for the converged cases and avoids unnecessary computation for cases that do not converge under the present fixed-point update. For days that do not meet the stopping criterion, the final submitted profile at $K_{\max}$ is used in the annual performance calculation, and convergence counts are reported separately.

\section{Conclusion}

This paper proposed a sharing coefficient-based demand response framework for renewable energy communities. Unlike conventional applications in which sharing coefficients are used after operation for surplus allocation and billing, the proposed framework uses the allocation outcome in the day-ahead stage to form household-specific price signals. The method links the economic signal received by each household to the composition of its allocated energy, distinguishing same-feeder sharing, inter-feeder sharing, and grid import.

The framework was developed for both feeder-aware and feeder-agnostic dynamic proportional sharing. In the feeder-aware case, surplus is first allocated within each feeder and only the residual surplus is allocated at the community level. In the feeder-agnostic case, the allocation is performed directly at the community level. The resulting prices are used in a household demand response problem that preserves daily energy consumption while allowing demand to shift across hours. Since the price signal depends on the submitted demand profiles and the submitted profiles change in response to the price, an iterative coordination procedure between the energy community manager and the households was used to obtain a consistent day-ahead schedule.

The numerical results show that the proposed allocation-based price signals can substantially reduce reverse power flow in low voltage feeders. Over the selected days with high reverse flow, average feeder reverse energy decreased from 44.21 kWh/day in the case without demand response to 24.32 kWh/day with feeder-aware demand response coordination and 24.51 kWh/day with feeder-agnostic coordination. The same cases reduced the average community bill by 21.18\% and 20.19\%, respectively. On the representative day of May 14, 2020, the feeder-aware strategy reduced total feeder reverse energy from 40.51 to 11.80 kWh and achieved the lower peak reverse power, main-grid export, network loss, and community bill compared to the case without demand response and the feeder-agnostic demand response case.

The annual evaluation confirms that the observed behavior is not limited to the selected days that have high reverse power. Over 345 complete days, annual feeder reverse energy decreased from 3618.4 kWh to 1120.2 kWh with feeder-aware coordination and to 1218.2 kWh with feeder-agnostic coordination. Also, exported energy to the main grid was strongly reduced from 2015.5 kWh to 163.4 kWh and 173.6 kWh, respectively. Additionally, the annual bill savings were lower than those obtained on the days with higher reverse energy flow because many days have less PV surplus and therefore less opportunity for demand shifting. Nevertheless, the feeder-aware strategy consistently achieved the lowest annual reverse flow energy, main-grid export, network loss, and community bill.

Overall, the results indicate that sharing coefficients can serve not only as settlement instruments but also as operational signals for demand response. The main improvement is obtained by using sharing-based prices to shift demand toward periods of local PV availability, while feeder-aware allocation provides an additional, moderate grid-oriented benefit by accounting for the location of households in the low voltage network.

\end{document}